\documentclass[twocolumn,showpacs,preprintnumbers,amsmath,amssymb]{revtex4}
\usepackage{graphicx}
\begin{document}

\title{Finite-temperature effects on the number fluctuation of ultracold atoms
across the Superfluid to Mott-insulator transition}

\author{Xiancong Lu and Yue Yu}
\affiliation{Institute of Theoretical Physics, Chinese
Academy of Sciences, P.O. Box 2735, Beijing 100080, China}
\date{\today}

\begin{abstract}
We study the thermodynamics of ultracold Bose atoms in optical
lattices by numerically diagonalizing the mean-field Hamiltonian
of the Bose-Hubbard model. This method well describes the behavior
of long-range correlations and therefore is valid deep in the
superfluid phase. For the homogeneous Bose-Hubbard model, we draw
the finite-temperature phase diagram and calculate the superfluid
density at unity filling. We evaluate the finite-temperature
effects in a recent experiment probing number fluctuation [Phys.
Rev. Lett. \textbf{96}, 090401 (2006)], and find that our
finite-temperature curves give a better fitting to the
experimental data, implying non-negligible temperature effects in
this experiment.
\end{abstract}


\maketitle

\section{Introduction}\label{1}

The ultracold Bose atoms in optical lattices have opened a new
window to investigate the strongly correlated systems with highly
tunable parameters \cite{Bloch}. The basic physics of these
ultracold atomic systems is captured by the Bose-Hubbard model,
whose most fundamental feature is the existence of superfluid to
Mott-insulator (MI) phase transition \cite{fisher,Jaksch}. In a
shallow optical lattice, the ultracold atoms are in superfluid
phase, which can be well-described by a macroscopic wave function
with long-range phase coherence \cite{gre}. In this case, phase
fluctuation vanishes and on-site number fluctuation diverges.
Whereas, in a deep optical lattice, the atoms are in MI phase if
the filling factor is fixed to an integer, leading to zero on-site
number fluctuation and divergent phase fluctuation
\cite{gre,Orzel}.

The physics of the MI phase is that, when the repulsive
interaction between atoms is large enough, the number fluctuation
would become energetically unfavorable, forcing the system into a
number-squeezed state. This interaction induced MI phase plays an
important role in the strongly correlated systems, as well as
various quantum information processing schemes \cite{Rabl}. In the
past, a series of ultracold-atom experiments have been performed
to detect this number-squeezed MI phase through the observation of
increased phase fluctuations \cite{Stoferle,gre,Orzel} or through
an increased time scale for phase diffusion
\cite{Greiner,F.Gerbier1}.

Recently, a continuous suppression of on-site number fluctuation
was directly observed by Gerbier \emph{et al.} by monitoring the
suppression of spin-changing collisions across the superfluid to
Mott-insulator transition \cite{F.Gerbier1}. By using a far
off-resonant microwave field, the spin oscillations for doubly
occupied sites can be tuned into resonance, and its amplitude is
directly related to the probability of finding atom pairs in a
lattice site. It was shown by Gerbier \emph{et al.} that, for a
small atom number, the oscillation amplitude is increasingly
suppressed when the lattice depth is increased and completely
vanishes for sufficiently large lattice depth. They also compared
the experimental results with the prediction of the Bose-Hubbard
model within a mean-field approximation at zero temperature.
However, the theoretical curves at zero temperature do not fully
fit with the experimental data, especially in low optical
lattices. An important reason for this deviation is the neglect of
finite-temperature effects in the calculation. This motivates us
to study the finite-temperature properties of these ultracold
atoms expecting a better fitting to the experimental data.

In the past, various theoretical approaches have been used to
investigate the zero temperature properties of the Bose-Hubbard
model such as the mean-field approximation
\cite{fisher,oosten,Sheshardi}, strong-coupling expansion
\cite{freericks}, Gutzwiller projection ansatz
\cite{Jaksch,Krauth,Schroll}, and quantum Monte Carlo simulation
\cite{QMC}. There are also a few works focusing on the
finite-temperature properties of the Bose-Hubbard model, such as
the standard basis operator method \cite{Sheshardi,Konabe}, slave
particle approach \cite{D,yy}, coarse-graining mean-field
approximation \cite{Kampf}, analytic investigation of fixed point
\cite{pv}, and some others \cite{DeMarco,finiteT}. However, most
of these approaches are based on the perturbation theory in terms
of small superfluid order parameter and therefore are only valid
near the superfluid to normal-liquid phase transition or in the
Mott region. Recently, Oosten \emph{et al.} have used an effective
approach, which is a finite-temperature extension of the zero
temperature mean field theory proposed by Sheshardi \emph{et al.}
\cite{Sheshardi}, to study the thermodynamics of the system with
large filling factor \cite{oo}. This mean-field theory correctly
includes the long-rang correlations \cite{Schroll,Garcia-Ripoll}
and thus is valid deep in the superfluid phase \cite{note}. In
this paper, we will use it to calculate the thermodynamic
quantities from the high superfluid region to the normal-liquid
(or MI) region.

This paper is organized as follows. In Sec. \ref{2}, we will
describe the finite-temperature mean-field theory for the
Bose-Hubbard model. We will draw the finite-temperature phase
diagram and investigate the system with unity filling factor. In
Sec. \ref{3}, we will calculate the finite-temperature number
fluctuations and compare them with the experimental results of
Ref. \cite{F.Gerbier1}. In Sec. \ref{4}, we will give our
conclusions.

\section{Finite-temperature mean-field theory to Bose-Hubbard model}\label{2}
We consider an ultracold Bose atom gas confined in a
three-dimensional optical lattice potential. In real experiments,
a slow varying trapping potential is superimposed onto the lattice
potential. We only pay attention to the homogeneous case in this
section and will consider the trapping potential in the next
section. The homogenous Bose atom system can be well-described by
the following Bose-Hubbard Hamiltonian \cite{Jaksch}:
\begin{eqnarray}\label{bhmodel}
H=-t\sum_{<ij>}b^\dag_ib_j-\mu\sum_in_i+\frac{U}{2}\sum_in_i(n_i-1).
\end{eqnarray}
Here $b_i^\dag$ is the creation operator at site $i$,
$n_i=b_i^\dag{b_i}$ is the particle number operator, and
$\langle{ij}\rangle$ denotes the sum over nearest neighbor sites.
$t$ and $U$ are the hopping amplitude and on-site interaction,
respectively. In the mean-field approximation \cite{Sheshardi},
the hopping term is decoupled as
\begin{eqnarray}
 b_i^\dag b_j&=&\langle b_i^\dag\rangle b_j
+\langle b_j^\dag\rangle b_i-\langle b_i^\dag\rangle\langle
b_j^\dag\rangle\nonumber\\
&=&\phi(b_i^\dag+b_j)-\phi^2,
\end{eqnarray}
where $\phi=\langle b_i^\dag\rangle=\langle b_i\rangle$ is the
superfluid order parameter. The Hamiltonian of the Bose-Hubbard
model can be written as a sum over single-site terms, $H=\sum_i
H_i$, where
\begin{eqnarray}
H_i=\frac{U}{2}n_i(n_i-1)-\mu n_i-zt\phi(b_i^\dag+b_i)+zt\phi^2,
\end{eqnarray}
with $z$ being the number of nearest neighbors. In Ref.
\cite{Sheshardi}, Sheshardi \emph{et al.} have studied the
zero-temperature properties of the Bose-Hubbard model by
diagonalizing the Hamiltonian $H_i$ in the occupation number basis
$\{|n\rangle\}$ truncated at a finite value $n_t$. Here, we will
extend this method to include the temperature effects \cite{oo}
and investigate the finite-temperature properties of the same
model. We first obtain the matrix of $H_i$ in the truncated
occupation number basis, which has a symmetric tridiagonal form:
\begin{eqnarray}
\left(\begin{array}{ccccccc}
d(1)&e(1)&0&\hdots&&&0\\
e(1)&d(2)&e(2)&&&&\vdots\\
0&e(2)&d(3)&&&&\\
\vdots&&&\vdots&&&\\
&&&&d(n_t-1)&e(n_t-1)&0\\
&&&&e(n_t-1)&d(n_t)&e(n_t)\\
0&\hdots&&&0&e(n_t)&d(n_t+1)
\end{array}\right),
\end{eqnarray}
where the diagonal elements $d(k)$ and subdiagonal elements $e(k)$
are
\begin{eqnarray}
d(k)=\frac{U}{2}(k-1)(k-2)-\mu(k-1)+zt\phi^2,
\end{eqnarray}
\begin{eqnarray}
e(k)=-\sqrt{k}zt\phi,
\end{eqnarray}
and all other matrix elements are zero. We diagonalize this matrix
to obtain the energy spectrum $\{E_k\}$ and eigenstates
$|\psi_k\rangle$, and then evaluate the partition function and the
free energy,
\begin{eqnarray}
Z=\sum_{k=1}^{n_t+1} e^{-\beta E_k},~~~
F=-\frac{1}{\beta}\ln{Z}.
\end{eqnarray}
For given $U$, $t$, $\mu$, and $T$, the superfluid order parameter
$\phi$ can be determined by minimizing the free energy, i.e.,
\begin{eqnarray}
\frac{\partial F}{\partial \phi}\biggr|_{U,t,\mu,T}=0.
\end{eqnarray}
The region with nonzero $\phi$ is identified as the superfluid
phase while the region with $\phi=0$ as the Mott-insulator or
normal-liquid phase. After determining $\phi$, it is easy to
calculate other physical quantities such as the superfluid density
$\rho_s$ and average density $\rho$ with
\begin{eqnarray}
\rho_s=\phi^2,~~~\rho=\langle n\rangle=\frac{{\rm Tr} (n
e^{-\beta H})}{Z}.
\end{eqnarray}

\begin{figure}
\begin{center}
\includegraphics[width=0.9\columnwidth]{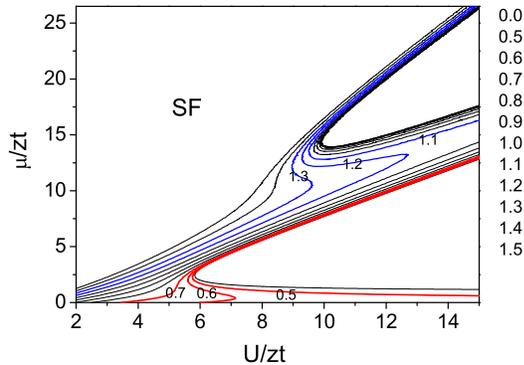}\\
\caption{\label{finiteT-muU}(Color online) The finite-temperature
phase diagram of the Bose-Hubbard model in the $\mu/zt$-$U/zt$
plane. The temperature is increasing when evolving away from the
zero-temperature Mott lobes. The temperature $T/zt$ of each curve
is listed in a column on the right side of the figure. The red
curves with values 0.5, 0.6, and 0.7 on them indicate the
crossover region for the first Mott lobe, and the blue ones with
1.1, 1.2, and 1.3 on them indicate the crossover for the second
lobe.}
\end{center}
\end{figure}

\begin{figure}
\begin{center}
\includegraphics[width=0.9\columnwidth]{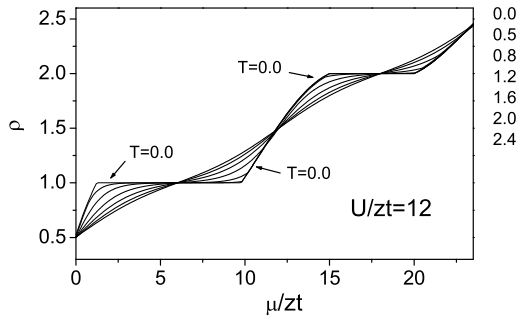}\\
\caption{\label{finiteT-rhomu} Average density $\rho$ as a
function of $\mu/zt$ at a fixed interaction $U/zt=12$. The
temperature is increasing when evolving away from the
zero-temperature curve, which is indicated by the arrows in the
figure. We list the temperature of each curve in the column on the
right side of the figure.}
\end{center}
\end{figure}

\begin{figure}
\begin{center}
\includegraphics[width=0.9\columnwidth]{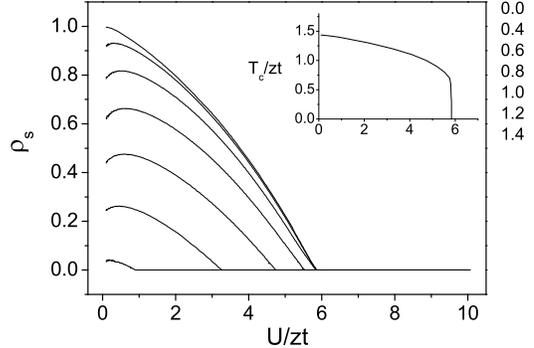}\\
\caption{\label{rhos} Superfluid density $\rho_s$ vs interaction
$U/zt$ for different temperatures in the case of integer filling
$\rho=1$. The temperature of each curve is increasing from up to
down, with their values listed in the column on the right side. In
the inset, the critical temperature $T_c/zt$ of the phase
transition is plotted as a function of interaction $U/zt$.}
\end{center}
\end{figure}

\begin{figure}
\begin{center}
\includegraphics[width=0.9\columnwidth]{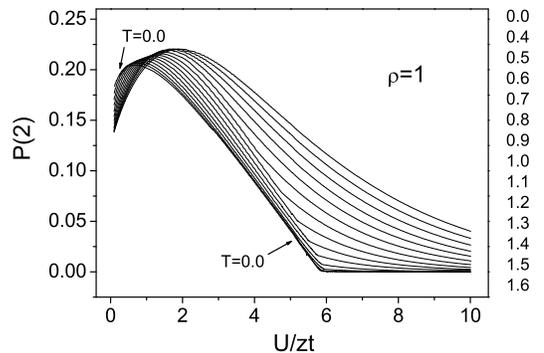}\\
\caption{\label{P2} Probability $P(2)$ of a site to be doubly
occupied vs interaction $U/zt$ for different temperatures, with
fixing filling factor $\rho=1$. The temperature of each curve is
decreasing from up to down, with their values listed on the right
side as usual.}
\end{center}
\end{figure}

We show our main results in Figs. \ref{finiteT-muU} - \ref{P2}.
The finite-temperature phase diagrams are plotted in Fig.
\ref{finiteT-muU} in the $\mu/zt$-$U/zt$ plane. The different
curves represent the phase boundaries between superfluid and
normal liquid (or MI) at different temperatures. One can see that
the zero-temperature Mott lobes are enlarged and disappear
gradually with the increasing temperature. This is the so-called
MI to normal-liquid crossover. We see that the crossover
temperature is about $0.5-0.7 zt$ around the first Mott lobe (red
curves), while it is about $1.1-1.3 zt$ around the second one
(blue curves). Note that we set $k_B=1$ throughout this work. The
second Mott lobe is more stable against the temperature than the
first because of the larger interaction $U$.

In Fig. \ref{finiteT-rhomu}, we show the average density $\rho$ as
a function of $\mu/zt$ for different temperatures and fixed
interaction $U/zt=12$. At zero temperature, there are plateaus on
the curve with integer filling factor and zero compressibility.
When increasing the temperature, these plateaus are shrunk and
eventually disappear. The qualitative behaviors of all these
curves are the same as those of previous finite-temperature works
based on perturbation theory \cite{Konabe,Kampf,D,yy}. However,
our results persist deep into the superfluid phase.

We then turn to investigate the system with an integer filling
factor, $\rho=1$, at which the superfluid to Mott-insulator phase
transition may occur. The evolution of superfluid density as a
function of interaction at different temperatures is shown in Fig.
\ref{rhos}. At zero temperature, $\rho_s$ drops to zero at a
critical interaction $U_c/zt=5.83$, in good agreement with the
analytical result \cite{oosten}. When increasing the temperature,
the superfluid density as well as the critical interaction is
reduced. Both of them vanish when $T/zt$ is larger than 1.45. We
record the critical interaction for every temperature, and plot
inversely the critical temperature $T_c/zt$ as a function of
interaction in the inset. This is consistent with our previous
result in Ref. \cite{yy}.

In Fig. \ref{P2}, we show the probability $P(2)$ of a site to be
doubly occupied as a function of interaction $U/zt$ for different
temperatures. At zero temperature, there is an obvious suppression
of $P(2)$ with the increasing interaction, i.e., $P(2)$ reduces to
zero when entering the MI region. However, one can see that the
suppression is weakened as the temperature is increased. This
phenomenon has recently been observed by Gerbier \emph{et al.} by
using a spin-changing collision technique \cite{F.Gerbier1}. We
will discuss this in detail in the next section.

\begin{figure}
\begin{center}
\includegraphics[width=0.9\columnwidth]{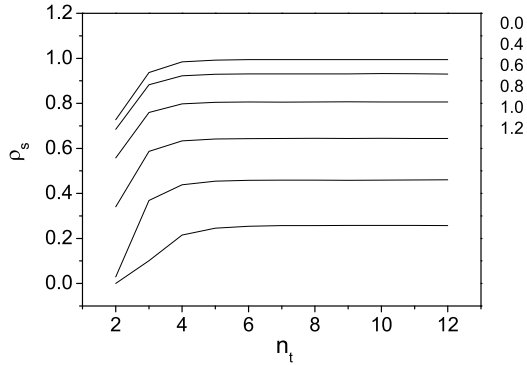}\\
\caption{\label{rhos_nt} Superfluid density $\rho_s$ vs truncation
$n_t$ for different temperatures. The density is $\rho=1$ and the
interaction is $U/zt=0.2$. The temperature of each curve is
increasing from up to down, with their values listed on the
right.}
\end{center}
\end{figure}

In the final of this section, we evaluate the effect of finite
truncation to the occupation number basis. As we know, the smaller
$U/zt$ the larger $n_t$ is needed for small error
\cite{Sheshardi}. We then plot in Fig. \ref{rhos_nt} the
superfluid densities $\rho_s$ of a unity filling system versus the
truncation $n_t$ for a very small interaction $U/zt=0.2$. One can
see that the finite-truncation approximation is quite worse in the
small $n_t$ region, i.e., the magnitude of $\rho_s$ with $n_t$ is
quite different from that with $n_t+1$. This finite-truncation
effect is strengthened when increasing the temperature. However,
when $n_t$ is larger than 6, it becomes very small even for a high
temperature. Typically, the difference of $\rho_s$ between
$n_t=10$ and $n_t=11$ is smaller than 0.5\%, which means the
effect of finite truncation is very small. Therefore we choose
$n_t=10$ in all of our calculations.

\section{The finite-temperature number fluctuations: comparing to experiment}\label{3}
In this section, we will evaluate the finite-temperature number
fluctuations of ultracold atoms in optical lattices and compare
them with the experimental data in Ref. \cite{F.Gerbier1}. The
ultracold atomic gas is confined in a three-dimensional optical
lattice potential $V_{OL}(\textbf{r})$ with
\begin{eqnarray}
V_{OL}(\textbf{r})=V_0\sum_{j=1}^3\sin^2(kr_j),
\end{eqnarray}
where $V_0$ is the lattice depth, $k=2\pi/\lambda$ is the wave
vector with $\lambda$ being the laser wavelength. In addition, the
atomic gas is also subjected to a trapping potential
$V_{ext}(\textbf{r})$, which can be approximately considered as a
harmonic one \cite{F.Gerbier2},
\begin{eqnarray}
V_{ext}(\textbf{r})=\frac{1}{2}m\omega^2\textbf{r}^2,
\end{eqnarray}
with the trapping frequency given by
\begin{eqnarray}
\omega=\sqrt{\omega_m^2+\frac{8V_0}{mw^2}}.
\end{eqnarray}
Here $m$ is the atom mass, $\omega_m$ is the frequency of the
magnetic trap where the condensate is initially formed, and $w$ is
the waist of the lattice beams \cite{F.Gerbier1,F.Gerbier2}. If
this trapping potential is very shallow and varies slowly across
the atomic cloud, we can treat it within the local density
approximation (for more details, see Ref. \cite{F.Gerbier2}).

We now specify the values of the parameters used in our
calculations. The hopping amplitude $t$ and on-site interaction
$U$ of the Bose-Hubbard model can be determined by
\begin{eqnarray}
t=\int{d\textbf{r}w^*(\textbf{r}-\textbf{r}_i)\left(-\frac{\hbar^2}{2m}\nabla^2+
V_{OL}(\textbf{r})\right)w(\textbf{r}-\textbf{r}_j)},
\end{eqnarray}
\begin{eqnarray}
U=\frac{4\pi a_s\hbar^2}{m}\int{d\textbf{r}|w(\textbf{r})|^4},
\end{eqnarray}
where $a_s$ is the s-wave scattering length and $w(\textbf{r})$ is
the Wannier function. Here, we use the approximate expressions
\cite{F.Gerbier2}
\begin{eqnarray}
\frac{t}{E_r}=1.43\left(\frac{V_0}{E_r}\right)^{0.98}
\exp{(-2.07\sqrt{V_0/E_r})},
\end{eqnarray}
\begin{eqnarray}
\frac{U}{E_r}=5.97\left(\frac{a_s}{\lambda}\right)\left(\frac{V_0}{E_r}\right)^{0.88},
\end{eqnarray}
where $E_r=h^2/2m\lambda^2$ is the single-photon recoil energy.
The triplet scattering length for the $^{87}$Rb atom is $a_s=5.61$
nm \cite{Pethick}, the laser wavelength $\lambda=842$ nm, the
lattice spacing $d=\lambda/2=421$ nm, the magnetic trapping
frequency $\omega_m=2\pi\times16$ Hz, and the waist $w=136$
$\mu$m. All these values coincide exactly with those in the real
experiment \cite{F.Gerbier1}.

\begin{figure}
\begin{center}
\includegraphics[width=0.9\columnwidth]{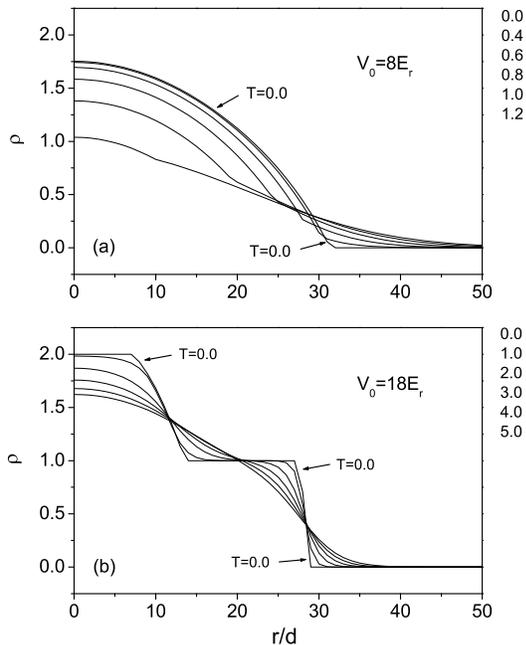}\\
\caption{\label{rho-r} Density distribution of ultracold atom gas
with a total atom number $N=1.0\times 10^5$ in different optical
lattice depths $V_0=8E_r$ (a) and $V_0=18E_r$ (b). The
zero-temperature curves are indicated by arrows in the figure. The
temperature of each curve is increasing when evolving away from
the zero-temperature one, with their values listed in the columns
on the right side.}
\end{center}
\end{figure}

We show in Fig. \ref{rho-r} the density distribution of ultracold
atomic gas with total number $N=1.0\times 10^5$ in two typical
optical lattice depths. We can see that, at zero-temperature and
in the low optical lattice [Fig. \ref{rho-r}(a)], the density
decreases smoothly when leaving the center of the trap. However, a
Mott-shell structure forms in the deep lattice [Fig.
\ref{rho-r}(b)]. When increasing the temperature, the atomic cloud
is expanding and the Mott-plateaus in the deep lattice are melting
gradually. Most recently, the formation of Mott-shell structure is
directly observed \cite{Folling,campbell}. The calculation of the
finite-temperature effects in these experiments is ongoing.

\begin{figure}
\begin{center}
\includegraphics[width=0.9\columnwidth]{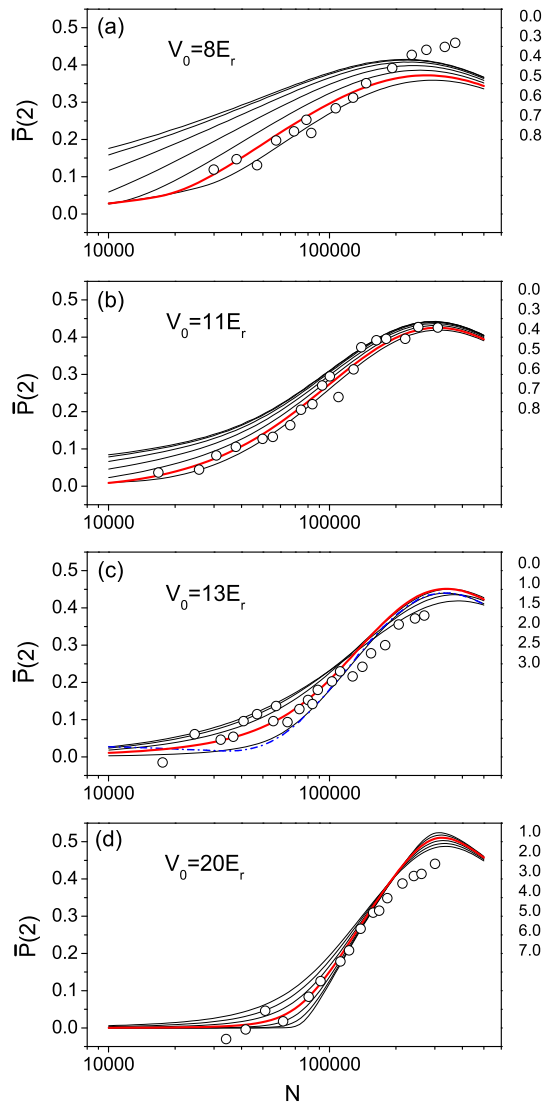}\\
\caption{\label{P2-N} (Color online) The probability $\bar{P}(2)$
of finding atoms in the doubly occupied sites, averaged over the
whole atomic cloud, vs the total atom number $N$ for different
lattice depths and temperatures. The  circles are the experimental
results reproduced from Ref. \cite{F.Gerbier1}. In panels (a) and
(b), the temperature of each curve is increasing from up to down.
In panel (c), the blue dash-dot curve corresponds to $T/zt=0.0$.
For other curves, the temperature is decreasing from up to down in
the small $N$ region, but is increasing from up to down in the
large $N$ region. Panel (d) has a similar temperature trend. From
up to down, the temperature is decreasing in the small $N$ region
but is increasing in the large $N$ region. As usual, the value of
temperature is listed on the right side of each panel. The
temperatures of the red curves in panels (a), (b), (c), and (d)
are $T/zt=$0.7, 0.7, 1.5, and 4.0, respectively. }
\end{center}
\end{figure}

We now turn to the number fluctuation experiment performed by
Gerbier \emph{et al.} \cite{F.Gerbier1}. In explaining their
experimental results, the authors used a zero-temperature
mean-field theory, which gives the correct trend. However, they
also found some discrepancies, especially in the low lattice
depths \cite{F.Gerbier1}. We find that these discrepancies can be
diminished by taking the finite-temperature effect into account.
We show the probability $\bar{P}(2)$ of finding atoms in the
doubly occupied sites, averaged over the whole atomic cloud, in
Fig. \ref{P2-N} as a function of total atom number $N$ for
different lattice depths and temperatures. The zero temperature
curve in each panel coincides with the theoretical estimate in
Ref. \cite{F.Gerbier1}. One can see that their deviation from the
experimental data, which are denoted by circles in the figure, is
quite obvious. In panel (a) where the atoms are in superfluid
phase, $\bar{P}(2)$ always decreases with the increasing
temperature. When comparing with the experimental data, we can see
that for small atom number the discrepancy between the
zero-temperature curve and the experimental data can be reduced by
taking the temperature into account. In panel (b), we observe that
the finite-temperature curve with a proper temperature, e.g., the
red curve with $T/zt=0.7$, agrees well with the experimental data
in the whole range of $N$. In panel (c) where the atoms are near
the superfluid to Mott-insulator transition, the influence of the
temperature is a little complex. In the small $N$ region,
$\bar{P}(2)$ first decreases and then increases with the
increasing temperature. The decrease at the beginning is due to
the suppression of superfluid density by the temperature, and the
following increase is caused by the temperature exciting more
atoms to occupy the eigenstate $|2\rangle$. In the large $N$
region, however, the behavior of $\bar{P}(2)$ is inverse: it first
increases and then decreases with the increasing temperature.
Again, we find that the curve with high temperature (e.g., the red
curve with $T/zt=1.5$) fits fairly well with the experimental
data. In panel (d) where the atoms are deep in the Mott region,
the amplitude of $\bar{P}(2)$ is improved in the small $N$ region,
but is suppressed in the large $N$ region, as the temperature
increases. This behavior is consistent with the experimental data
too. The investigation of these finite-temperature effects has its
own interest, for it provides a possible way to estimate the
temperature of the atomic gas \cite{Folling}. In our case, $1.0zt$
yields several ten nano-Kelvin and the temperature regime where
the experimental data located in is $10^1\sim10^2$ nK, therefore
our estimate to temperature is quite reasonable.

Before closing this section, we observe that, in the large $N$
region, there are still some inconsistencies between our results
and the experimental data. Within the deep superfluid regime [Fig.
\ref{P2-N}(a)], $\bar P(2)$ is always smaller than the
experimental data in the large $N$ region. A possible reason for
this discrepancy is the large probability for triplet occupation,
which is due to the high density in the center of the trap
($\rho\approx3$) and the large number fluctuation of the system in
so low lattice potential. In experiment, the presence of large
triplet occupation may change the spin resonance condition,
resulting in a considerable contribution of spin oscillations on
triply occupied sites to the observed oscillation amplitude
\cite{F.Gerbier1,Capogrosso-Sansone}. This makes the experimental
measurements larger than the calculated $\bar P(2)$ (see also Fig.
4 of Ref. \cite{F.Gerbier1}). Within the deep MI regime [Fig.
\ref{P2-N}(d)], our finite-temperature consideration does give a
correct trend in the large $N$ region. However, the reason for the
further suppression of oscillation amplitude is not uncovered yet,
which cannot be simply attributed to the temperature effects.

\section{Conclusions}\label{4}

We studied the finite-temperature properties of the Bose-Hubbard
model by numerically diagonalizing the mean-field Hamiltonian on a
truncated occupation number basis. This method can give the
thermodynamics deep in the superfluid phase. For the homogenous
Bose-Hubbard model, we calculated the finite-temperature phase
diagram and plotted the density versus chemical potential curve.
We demonstrated the evolution of zero-temperature Mott lobes and
Mott plateaus when the temperature is increasing. We then
specially investigated the system with unity filling factor. The
superfluid density and the probability $P(2)$ of finding a site to
be doubly occupied were plotted as a function of interaction. We
found that the superfluid density is reduced and the suppression
of $P(2)$ is weakened as the temperature increases. In the second
part of this paper, we evaluated the finite-temperature effects in
a recent experiment probing the number fluctuation
\cite{F.Gerbier1}. We showed that the calculated
finite-temperature curves fit the experimental data better than
the zero-temperature ones. This implies that the
finite-temperature effects in this experiment are quite large.

\begin{acknowledgments}
We thank Fabrice Gerbier for a lot of useful discussions and
explaining their experimental and theoretical results. We are also
grateful to Jingyu Gan, Liang He, Shijie Hu, Yuchuan Wen, Shijie
Yang, and Hao Yin for helpful discussions. This work was supported
in part by Chinese National Natural Science Foundation.
\end{acknowledgments}


\begin{references}
\bibitem{note} Note that, although the mean field theory correctly
includes the long-rang correlations, it does fail to describe the
short-rang correlations between different lattice sites, which,
however, can be obtained by the perturbation theory
\cite{Schroll,F.Gerbier2}. Also note that this comparison is
performed for the homogeneous system.
\end{references}
\end{document}